\documentclass[3p]{elsarticle}

\usepackage{amsmath,amscd}
\usepackage{graphicx}%
\usepackage{subfig} %
\usepackage{verbatim}
\usepackage{dcolumn}%
\usepackage{bm}%
\usepackage{adjustbox}
\usepackage{float}

\newcommand{\R}{{\mathbf R}}
\newcommand{\n}{{\mathbf n}}
\newcommand{\rr}{{\mathbf r}}
\newcommand{\e}{{\mathbf e}}
 
\begin{document}

\title{Computation of the magnetostatic interaction between linearly magnetized polyhedrons}

\author[soton]{Dmitri Chernyshenko}%
\author[soton]{Hans Fangohr}%
\address[soton]{%
Engineering and the Environment,
University of Southampton, Southampton SO17 1BJ,
United Kingdom
}%

\date{\today}

\begin{abstract}
In this paper we present a method to accurately compute the energy of the
magnetostatic interaction between linearly (or uniformly, as a special case) magnetized polyhedrons. The method has applications in finite element micromagnetics, or more generally in computing the magnetostatic interaction when the magnetization is represented using the finite element method (FEM).
 
The magnetostatic energy is described by a six-fold integral that is singular when the interaction regions overlap, making direct numerical evaluation problematic. To resolve the singularity, we evaluate four of the six iterated integrals analytically resulting in a 2d integral over the surface of a polyhedron, which is nonsingular and can be integrated numerically. This provides a more accurate and efficient way of computing the magnetostatic energy integral compared to existing approaches.

The method was developed to facilitate the evaluation of the demagnetizing interaction between neighouring elements in finite-element micromagnetics and provides a possibility to compute the demagnetizing field using efficient fast multipole or tree code algorithms.

 \end{abstract}

\maketitle

\date{\today}

\section{Introduction}
In the continuum form of the Landau-Lifshitz-Gilbert equation, the effective field $\mathbf H_\mathrm{eff}(\mathbf r)$ is the functional derivative of the total energy functional $E(\mathbf M)$ with respect to the magnetization $\mathbf M(\mathbf r)$~\cite{Brown1963}: 
\begin{gather}
\mathbf H(\mathbf r) = - \frac{1}{\mu_0} \frac{\delta E}{\delta \mathbf M(\mathbf r)} \label{eq-llg} \\
E = E_\text{Zeeman} + E_\text{demag} + E_\text{anisotropy} + E_\text{exchange} + \ldots
\end{gather}

In numerical micromagnetics, the dynamics of magnetization are described by the semi-discretized Landau-Lifshitz-Gilbert equation, in which the motion of magnetization is computed from the discretized effective field. It is desireable to preserve the relation~\eqref{eq-llg} between effective field and total energy in the semi-discretized formulation~\cite{Miltat2007} --- if~\eqref{eq-llg} holds for the discrete system, then the total energy will decrease in the simulation, simplifying the use of energy-based criteria for the control of the simulation or the search for an equilibrium.  

In order to preserve~\eqref{eq-llg}, the effective field has to be computed from the discretized total energy function using the corresponding discrete counterpart to the functional derivative~$\delta E/\delta \mathbf M(\mathbf r)$. For the exchange, anisotropy, and Zeeman terms this is usually straightforward, however for the demagnetizing field it is more difficult. In finite difference (FD) micromagnetics, it can be achieved by computing the total demagnetizing energy of the system using the analytical expression~\cite{Schabes1987,Newell1993,maicus1998magnetostatic,fukushima1998volume} for the demagnetizing tensor, and then differentiating with respect to the degrees of freedom~\cite{oommf,Miltat2007}. %
However, in finite element (FE) micromagnetics, the demagnetizing field is usually computed using the FEM/BEM method~\cite{Fredkin1990,Garcia-Cervera2006} where the field is derived from the magnetostatic potential, and the energy is not computed exactly. %

It is therefore desireable to be able to accurately and efficiently compute the total magnetostatic energy of a system represented by a set of polyhedral elements, with magnetization linear inside each polyhedron (as in the FE method with linear Lagrange elements). The total energy of the system is the sum of pairwise interactions between the polyhedrons and in this paper we describe how to compute this pairwise interaction. 

Given two interacting magnetized polyhedrons $\tau$  and $\tau'$ with arbitrary magnetizations $\mathbf M(\mathbf r)$ and $\mathbf M'(\mathbf r)$, the energy $E_{\tau\leftrightarrow\tau'}$ of the magnetostatic interaction between them is:
\begin{equation}
E_{\tau\leftrightarrow\tau'} = - \mu_0 \int_{\tau} \mathbf M(\mathbf r) \cdot \mathbf H'_{\mathrm{demag}}(\mathbf r) \, d\mathbf r
= \frac{\mu_0}{4\pi} \int_{\tau} \int_{\tau'} 
	\mathbf M(\mathbf r) \cdot
		\big(\nabla_{\mathbf r} \nabla_{\mathbf r'} \frac{1}{|\mathbf r-\mathbf r'|}\big) 
	\cdot \mathbf M'(\mathbf r') 
	\, d\mathbf r \, d \mathbf r'
\label{eq-pairwise}
\end{equation}
where  $H'_{\mathrm{demag}}(\mathbf r) = - \frac{1}{4 \pi} \nabla_{\mathbf r} \int_{\tau'} \mathbf M'(\mathbf r') \cdot \nabla_{\mathbf r'} (1/|\mathbf r-\mathbf r'|) \, d\mathbf r'$ is the demagnetizing (stray) field produced by the polyhedron~$\tau'$.

The straightforward approach of numerically computing the integral~\eqref{eq-pairwise} is problematic because it requires explicit integration over a 6-dimensional region of space; additionally, when the polyhedrons overlap or coincide, the integrand is singular and regular integration methods cannot be applied. 
Many analytical results are available for similar 3-fold integrals arising during the calculation of the field~\cite{Wilton1984,Graglia1987,Graglia1993,Fabbri2008,Fabbri2009}. However, none of the formulas can be easily adapted to this more complex 6d case. In~\cite{Masserey2005} a method is developed for removing the singularity in~\eqref{eq-pairwise} that can be applied to the uniform magnetization case but does not generalize easily to the linear magnetization case. A Fourier-transform method has been devised for the computation of magnetostatic energy~\cite{Beleggia2005,Beleggia2004}, however for the case investigated in this paper, the required 3d numerical integration in the Fourier space is still somewhat impractical. 

The rest of the paper describes the proposed method for the computation of this integral. The main approach is to analytically perform 4 out of 6 iterated integrals resulting in a 2d surface integral that is nonsingular and can be evaluated numerically using standard methods. This semi-analytical approach is similar to~\cite{Arcioni1997}; the use of notation and vector analysis in the analytical derivation is similar to the techniques in~\cite{Fabbri2008,Fabbri2009}.  

\section{Formulation of the problem}

For the purposes of computation, an arbitrary linear vector-valued function in space $\mathbf M(\mathbf r)$ can be represented by a $3 \times 4$ matrix $|| M_{ij} ||$: 
$\mathbf M(\mathbf r) = || M_{ij} || \cdot (1, r_x, r_y, r_z)^T$. However, performing analytical calculations for this general case is quite inconvenient; instead we only consider vector-valued linear functions of the form $A(\mathbf r) \mathbf M$ where $A(\mathbf r)$ is a scalar linear function and $\mathbf M$ is a constant vector. For the common case of a tetrahedral element, an arbitrary linear vector-valued function $\mathbf M(\mathbf r)$ can be reconstructed from the vertex values $\mathbf M_i$, $i=1\dots 4$: 
$\mathbf M(\mathbf r) = \sum_{i=1}^4 A_i(r) \mathbf M_i $, where $A_i(r)$ are the shape functions of the tetrahedron. 

\begin{figure} \centering
\includegraphics[width=10cm]{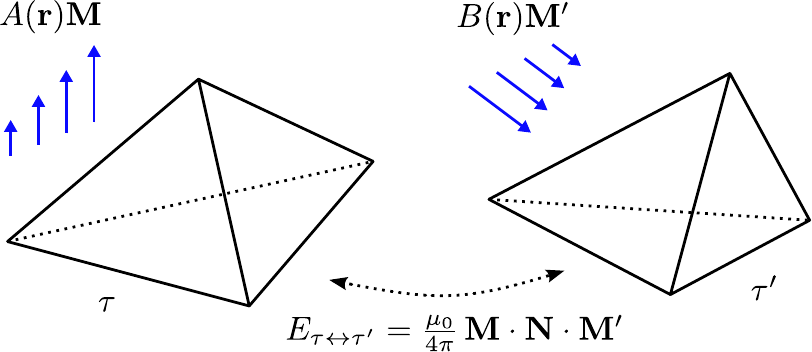}%
\caption{The energy $E_{\tau\leftrightarrow\tau'}$ of the magnetostatic interaction between  polyhedrons $\tau$  and $\tau'$ with magnetizations $A(\mathbf r) \mathbf M$ and $B(\mathbf r) \mathbf M'$, where $\mathbf M$ and $\mathbf M'$ are constant magnetization vectors and $A(\mathbf r)$, $B(\mathbf r)$ are dimensionless linear functions in space.}%
\label{fig-tets} 
\end{figure}

We perform the computations for a pair of interacting linearly magnetized polyhedrons $\tau$  and $\tau'$ with magnetizations $A(\mathbf r) \mathbf M$ and $B(\mathbf r) \mathbf M'$, where $\mathbf M$ and $\mathbf M'$ are constant magnetization vectors and $A(\mathbf r)$, $B(\mathbf r)$ are dimensionless linear functions in space (Figure~\ref{fig-tets}). From~\eqref{eq-pairwise}, the energy $E_{\tau\leftrightarrow\tau'}$ of the magnetostatic interaction between the polyhedrons is $E_{\tau\leftrightarrow\tau'} = \frac{\mu_0}{4 \pi} \, \mathbf M\cdot \mathbf N \cdot \mathbf M'$, where $\mathbf N$ is the symmetric $3\times3$ ``demagnetizing tensor'' 
\begin{equation}
\mathbf N = \mathbf N(A, B, \tau, \tau') = \int_\tau \int_{\tau'} A(\mathbf r) B(\mathbf r') \nabla_{\mathbf r} \nabla_{\mathbf r'} \frac{1}{|\mathbf r-\mathbf r'|}\, d\mathbf r \, d \mathbf r'
\label{eq-tettet}
\end{equation}

The goal of this paper is to compute this sixfold integral given the coordinates of the vertices of $\tau$ and~$\tau'$ and the coefficients of the linear functions $A(\mathbf r)$ and $B(\mathbf r)$. When the polyhedrons $\tau$ and $\tau'$  are separated, the integral can be computed numerically, however when the polyhedrons overlap or coincide, the integrand is singular and standard numerical integration is inaccurate. 

To deal with this issue, we analytically reduce the double volume integral~\eqref{eq-tettet} to a double surface integral, then evaluate the surface integral over $\mathbf r$ analytically, and the second  surface integral integral over $\mathbf r'$ numerically. This procedure is similar to the one employed in~\cite{Arcioni1997} --- the four analytical steps result in a surface integral with a bounded integrand that can be integrated numerically with reasonable accuracy and efficiency.

A short notice on units: the demagnetizing tensor commonly used in finite difference micromagnetics~\cite{Newell1993} is dimensionless, however the tensor~$\mathbf N$ computed in this paper~\eqref{eq-tettet} has units of volume.

\section{Method}

The analytical derivation proceeds in the following three steps:
\begin{itemize}
  \item transform the double volume integral~\eqref{eq-tettet} to a double surface integral using Gauss's theorem, removing linear factors via integration by parts 
  (Section~\ref{sec:vol2surface}), 
  \item express the integrand for the outer surface integral over $\mathbf r'$ as a linear combination of primitive terms  (Section~\ref{sec:surf2terms}), 
 \begin{align}
I_0(\tau; \mathbf r') &= \int_{\partial\tau - \mathbf r'} \frac{1}{|\R|} ds \label{eq-i0} \\ 
\mathbf I_1(\tau; \mathbf r') &= \int_{\partial\tau - \mathbf r'} \frac{\R}{|\R|} ds \\ 
\mathbf I_2(\tau; \mathbf r') &= \int_{\partial\tau - \mathbf r'} \frac{\R^{\otimes 2}}{|\R|} \label{eq-i2} ds
 \end{align}
 where $\partial \tau$ is the polygonal surface of the polyhedron $\tau$, $\partial \tau - \mathbf r'$ is the same surface shifted by $\mathbf r'$ according to the the substitution $\R = \mathbf r - \mathbf r'$, and $\otimes$ denotes tensor multiplication (i.e. $\mathbf R^{\otimes 2}$ is a symmetric tensor of rank 2).
   \item analytically integrate the primitive terms $\mathbf I_k$ over each polygonal surface, again by applying integration by parts, Stokes' theorem (for integration over a surface), and gradient theorem (for integration over a line), Section~\ref{sec:eval-ik}.
\end{itemize}

The equation \eqref{eq-surface-i} derived in step 2 (see Section~\ref{sec:analyticalresult}) together with the corresponding analytical formulas for the primitive terms $\mathbf I_k$ form the main result of the paper. 

\subsection{Auxiliary functions}

The derivation of the analytical formulas for $\mathbf I_k$ requires the computation of a number of auxiliary functions:

\begin{align}
J_0(\tau; \mathbf r') &= \int_{\partial \tau - \mathbf r'} |\R| ds \\ 
\eta_0(\R_1, \R_2) &= \int_{\R_1}^{\R_2} \frac{1}{|\mathbf R|} dl \\ 
\boldsymbol\eta_1(\R_1, \R_2) &= \int_{\R_1}^{\R_2} \frac{\R}{|\R|} dl \\ 
\lambda_0(\R_1, \R_2) &= \int_{\R_1}^{\R_2} {|\R|} dl \\ 
\boldsymbol\lambda_1(\R_1, \R_2) &= \int_{\R_1}^{\R_2} \R{|\R|} dl \\ 
 \end{align}
 
 The relation between the formulas for these functions is shown below (an arrow indicates that the formula at the source depends on the formula at the target):

\begin{equation}
\label{eq:formula-relations}
\begin{CD}
\eta_0 @<<< I_0 \\
@AAA @AAA \\  
\lambda_0 @<<< \mathbf I_1 \\ 
@AAA @AAA \\  
J_0 @<<< \mathbf I_2 @>>> \boldsymbol\lambda_1 @>>> \boldsymbol\eta_1 \\ 
\end{CD}
\end{equation}

\section{Analytical derivation --- reduction to a surface integral}
\label{sec:vol2surface}      

The first step is the conversion of~\eqref{eq-tettet} to a double surface integral.
Due to the high complexity of intermediate expressions the derivation was performed using a computer algebra system; the source code is included in the supplementary information for this paper~\cite{githubrepo}. Table~\ref{tab-vol-surf-conversion} shows the identities used in the derivation, they are applied repeatedly in a straightforward manner to integrate the terms and simplify the result. We chose to express the vector/tensor identities using tensor notation which is compact and more suitable for computer algebra than using explicit indices (with Einstein summation). 

After repeatedly applying the identities in Table~\ref{tab-vol-surf-conversion}, we arrive at the following formula for the demagnetizing tensor
\begin{equation}
\begin{split}
\mathbf N =
\int_{\partial \tau'} \int_{\partial \tau}  \Big[
\frac 12 A(\rr) (\n_{\rr'} \cdot (\rr - \rr')) \, \n_\rr \otimes\nabla B
- \frac 12 B(\rr') (\n_{\rr} \cdot (\rr - \rr')) \, \n_{\rr'} \otimes\nabla A
\\
+ \frac 16 (\n_{\rr} \cdot (\rr - \rr')) (\n_{\rr'} \cdot (\rr - \rr')) \, \nabla A \otimes\nabla B
+ A(\rr) B(\rr') \, \n_{\rr} \otimes \n_{\rr'}
\Big]
\frac{1}{|\rr - \rr'|}
ds\, ds'
\end{split}
\label{eq-surfsurf}
\end{equation}
where $\n_{\rr}$, $\n_{\rr'}$ are the normal vectors to the corresponding surfaces and $\nabla A$, $\nabla B$ are the (constant) gradient vectors for the linear functions $A(\mathbf r)$ and $B(\mathbf r)$. As expected, the formula is symmetrical under replacement $A \leftrightarrow B$, $\rr \leftrightarrow \rr'$ and reduces to Gauss's theorem when $A(\mathbf r)$ and $B(\mathbf r)$ are constant.

Side note: in principle, all derivations could be performed with scalars instead of tensors by computing the scalar counterpart to the integral~\eqref{eq-tettet}
\begin{equation}
N(\mathbf m, \mathbf m') = \mathbf m \cdot \mathbf N \cdot \mathbf m' = \int_\tau \int_{\tau'} A(\mathbf r) B(\mathbf r') 
(\mathbf m \cdot \nabla_{\mathbf r}) (\mathbf m' \cdot \nabla_{\mathbf r'}) \frac{1}{|\mathbf r-\mathbf r'|} \, d\mathbf r \, d\mathbf r'
 \end{equation}
With this method the auxiliary vectors $\mathbf m$ and $\mathbf m'$ would have to be included in all intermediate derivations;  
this would remove the need to keep track of tensor indices at the cost of slightly expanded notation.

\newcommand{\rot}{\operatorname{rot}}

\begin{table*}[htb]%
\centering
\begin{tabular}{|c|c|}%
\hline
Gauss's theorem &
\marginbox{2pt}{$ \displaystyle
\int_V \nabla_{\mathbf r} \, \mathbf F \, dr = \int_{\partial V} \n \otimes \mathbf F \, dr
$}%
\\
\hline
gradient of a product &
\marginbox{2pt}{$ \displaystyle
a \, \nabla_\mathbf r \mathbf F =
\nabla_\mathbf r [a \, \mathbf F]
- \nabla a \otimes \mathbf F
$}%
\\
\hline
integration of $\frac{(\rr-\rr')^{\otimes k}}{|\rr-\rr'|}$ &
\marginbox{2pt}{$ \displaystyle
\frac{(\rr-\rr')^{\otimes k}}{|\rr-\rr'|} =
\frac{1}{k+2} \nabla_\rr \cdot \frac{(\rr-\rr')^{\otimes \, k+1}}{|\rr-\rr'|}
$}%
\\
\hline
\end{tabular}%
\caption{Identities used in the conversion of the double volume integral~\eqref{eq-tettet} to a double surface integral. Here $\mathbf F(\rr)$ is a tensor of any rank and $a(\rr)$ is a scalar. When the gradient operator $\nabla$ is applied to a vector or tensor, we assume that the new tensor index is added at the front (i.e. $\nabla f_{jk} = \partial_i f_{jk}$). Similarly, divergence applies to the first index of a tensor (i.e. $\nabla f_{jk} = \partial_j f_{jk}$).} %
\label{tab-vol-surf-conversion}%
\end{table*}

\section{Analytical result --- integrand and the primitive terms~$\mathbf I_k$}
\label{sec:analyticalresult}  
\label{sec:surf2terms}      

The next step is to express the inner integral over~$\rr$ in~\eqref{eq-surfsurf} in terms of the primitive integrals~$\mathbf I_k$~\eqref{eq-i0}--\eqref{eq-i2}. By using the identity $A(\rr) = A(\rr') - (\rr - \rr')\cdot \nabla A$, we get

\begin{equation}
\begin{split}
\mathbf N = \int_{\partial \tau'} \Big[
\frac12 A(\mathbf r') (\mathbf n' \cdot \mathbf I_1) (\mathbf n \otimes \nabla B)
+ B(\mathbf r') (\nabla A \cdot \mathbf I_1)(\mathbf n \otimes \mathbf n')
- \frac12 B(\mathbf r') (\mathbf n \cdot \mathbf I_1) (\nabla A \otimes \mathbf n') \\
+ A(\mathbf r') B(\mathbf r') I_0 \, \mathbf n \otimes \mathbf n'
- \frac16 I_2(\mathbf n, \mathbf n') (\nabla A \otimes \nabla B)
+ \frac12 I_2(\nabla A, \mathbf n') (\mathbf n \otimes \nabla B)
\Big] ds'
\end{split}
\label{eq-surface-i}
\end{equation}

Note that the primitive integral $\mathbf I_k$ is a tensor of rank~$k$ and is a function of $\mathbf r'$.

\subsection{Evaluation of $\mathbf I_k$}
\label{sec:eval-ik}

The integral $I_0$ has been computed in~\cite{Fabbri2008} (eq.~(17) for $W_f(\mathbf r)$ in~\cite{Fabbri2008})
\begin{equation}
I_0 = \sum_{F\in\partial\tau - \mathbf r'}\left[ -\sum_{(\R_1, \R_2) \in \partial F} 
(\n \times \mathbf u \cdot \mathbf R_1) \, \eta_0(\R_1, \R_2) - 
(\n \cdot \R_f) \, \Omega(F) \right] 
\end{equation} 
where the outer sum is over the facets~$F$ of the polyhedron surface~$\partial\tau - \mathbf r'$, the inner sum is over the line edges $(\R_1, \R_2)$ of the facet, 
$\n$ is the facet normal, $\mathbf u = (\R_2 - \R_1)$ $/|\R_2-\R_1|$ is the unit vector along the edge, $\R_f$ is any point on the facet, and $\Omega(F))$ is the solid angle of the facet~$F$ from the origin. 	 

The integral $\mathbf I_1$ has been computed in~\cite{Fabbri2009} (eq.~(27) in~\cite{Fabbri2009})
\begin{equation}
\mathbf I_1 =
\sum_{F\in\partial\tau - \mathbf r'}\left[ 
- \sum_{(\R_1, \R_2) \in \partial F} (\n \times \mathbf u) \, \lambda_0(\R_1, \R_2) + (\n \cdot \R_f) \, \n  \, I_0(F)
\right]    
\end{equation}
Note that in the expression for $\mathbf I_1$, each term of the sum over the facets references the integral~$I_0(F)$ applied only to that facet, not the whole surface. 

The derivation of the formula for the integral $\mathbf I_2$ is shown in~\ref{sec:app_i2}
\begin{equation}
\mathbf I_2 = 
\sum_{F\in\partial\tau - \mathbf r'}\left[ 
	- \sum_{(\R_1, \R_2) \in \partial F} (\n \times \mathbf u) \otimes \lambda_1(\R_1, \R_2)  
	+ (\n^2 - \mathbf{Id}) \, J_0(F)
	+ (\n \cdot \R_f) \, \n \otimes \mathbf I_1(F)
\right]    
\end{equation}
Again, each term of the sum over the facets references the integrals $J_0(F)$ and $\mathbf I_1(F)$ applied only to that facet. 

The auxiliary function $\eta_0$ (used in the equations for $I_0$ and $\lambda_1$) has been computed in~\cite{Fabbri2008} (eq.~(18) for $w_e(\mathbf r)$)
\begin{equation}
\eta_0(\R_1, \R_2) = \ln \frac{|\R_1| + |\R_2| + |\R_2 - \R_1|}{|\R_2| + |\R_1| - |\R_2 - \R_1|}
`\end{equation} 

The auxiliary function $\lambda_0$ (used in the equations for $\mathbf I_1$ and $J_0$) has been computed in~\cite{Fabbri2009} (eq.~(22) for $\lambda_e(\mathbf r)$)
\begin{equation}
\lambda_0(\R_1, \R_2) = 
\frac12 \mathbf u (\R_2|\R_2| -  \R_1|\R_1|) + 
\frac12 |\R_1 \times \mathbf u|^2 \, \eta_0(\mathbf R_1, \mathbf R_2)
 \end{equation} 

The auxiliary function $J_0$ (used in the equation for $\mathbf I_2$) has been computed in~\cite{Fabbri2008} (eq.~(22) for $\Lambda_f(\mathbf r)$)
\begin{equation}
J_0 = \sum_{F\in\partial\tau - \mathbf r'}\left[ 
\frac13 \sum_{(\R_1, \R_2) \in \partial F} 
\mathbf n \times R_1 \cdot \mathbf u \, \lambda_0(\R_1, \R_2)
+ \frac13 (\R_f \cdot \n)^2 I_0  
\right]  
\end{equation}

The derivation of the formula for $\boldsymbol\eta_1$ (used in the equation for $\boldsymbol\lambda_1$) is shown in~\ref{sec:app_eta1} 
\begin{equation}
\boldsymbol\eta_1(\R_1, \R_2) = 
 	\mathbf u(|\R_2|  - |\R_1|)
 	- \mathbf u \times (\mathbf u \times \R_1) \, \eta_0(\R_1, \R_2) 
 \end{equation}

The derivation of the formula for $\boldsymbol\lambda_1$ (used in the equation for $\mathbf I_2$) is shown in~\ref{sec:app_lambda1} 
\begin{equation}
\boldsymbol\lambda_1(\R_1, \R_2) =
\Big(\frac12\mathbf{Id} + \frac16 \mathbf u \, (\mathbf u \cdot{})\Big)
 \Big[
	|\R| (\mathbf u \cdot \R) (\R  - \frac12 \mathbf u \, (\mathbf u \cdot \R))  \, \Big|_{\R_1}^{\R_2}
	+|\mathbf u \times \R_1|^2  (\mathbf{Id}  - \frac12 \mathbf u \, (\mathbf u \cdot{})) \, \boldsymbol\eta_1(\R_1, \R_2)
 \Big]
\label{eq-last}
\end{equation}
with the notation $f(\R)\big|_{\R_1}^{\R_2} \equiv f(\R_2) - f(\R_1) $.

\section{Numerical results}

In order to numerically verify the analytical results in Section~\ref{sec:analyticalresult} we need a way to compute the energy integral~\eqref{eq-pairwise} exactly or with sufficient precision. For a cuboid, we could do this if the magnetization was constant by using the analytical expression for the demagnetizing tensor~\cite{Schabes1987,Newell1993,maicus1998magnetostatic,fukushima1998volume}. However, for a constant magnetization the gradient terms $\nabla A$ and $\nabla B$ in our analytical expressions would be zero, and for a more comprehensive test we have to cover the case of (nontrivial) linear magnetization. The authors are not aware of an analytical result that could be used as a reference in this case, instead, as a reference we used a series of finite difference micromagnetic simulations with progressively increasing mesh size until convergence was reached.

\subsection{Test problem formulation}

The test system is a magnetized cuboid with dimensions $1.7 l \times 1.3 l \, \times l$, where $l$ is a arbitrary length parameter. A linear magnetization function $\mathbf M(\mathbf r)$ of the cuboid can be written in the general form  
$$\mathbf M(\mathbf r) = \mathbf L \cdot
\begin{pmatrix}
l\\r_x\\r_y\\r_z
\end{pmatrix}
$$
where $\mathbf L$ is a $3\times4$ matrix. In this equation we have used the length parameter $l$ to make the units of $\mathbf L$ uniform. 

The 12 entries of the matrix $\mathbf L$ can also be written as a column vector $\mathbf L_\mathrm{vec}$; we assume $\mathbf L_\mathrm{vec}$ is formed by stacking the columns of $\mathbf L$ (i.e, the first 3 entries of $\mathbf L_\mathrm{vec}$ come from the first column of $\mathbf L$, and so on). 

The demagnetizing energy $E = E(\mathbf L)$ of the cuboid is a quadratic function of $\mathbf L$ which we write in the form
$$
E(L) = -\frac12 \mu_0 l^5 \, \mathbf L_\mathrm{vec} \cdot \mathcal{E} \cdot \mathbf L_\mathrm{vec}
$$   

Here, $\mathcal{E}$ is a symmetric $12 \times 12$ matrix; the factor $l^5$ is introduced to make $\mathcal{E}$ dimensionless. The entries in $\mathcal{E}$ do not depend on $l$ but only on the aspect ratio of the cuboid ($1.7 \times 1.3 \times 1$ in our case).  

To verify our analytical formula, we compute the entries of the matrix $\mathcal E$ in two ways:

\begin{itemize}
  \item Via finite difference micromagnetic simulations with a progressively finer mesh until convergence, producing the reference matrix $\mathcal E_\mathrm{ref}$.
  \item By subdividing the cuboid into tetrahedrons and calculating pairwise interactions between the tetrahedrons using our new analytical formula~\eqref{eq-surface-i}--\eqref{eq-last}, producing the test matrix $\mathcal E_\mathrm{test}$.  
\end{itemize}

To estimate the error of computing $\mathcal E_\mathrm{test}$, we compute the relative error
$$
\eta = \frac{||\mathcal E_\mathrm{test}-\mathcal E_\mathrm{ref}||}{||\mathcal E_\mathrm{ref}||}
$$
using the sum-of-squares matrix norm
$$
||\mathcal{E}|| = \sum_{i,j} \varepsilon_{ij}^2 
$$

\subsection{Test results}

The reference matrix $\mathcal E_\mathrm{ref}$ was computed by performing finite diference simulations with mesh sizes $8\times 8 \times 8$, $16\times 16 \times 16$, $\ldots$, up to $128\times 128 \times 128$, and computing the Richardson's extrapolation estimate using the last two steps. The estimated relative error of computing  $\mathcal E_\mathrm{ref}$ (compared to the unknown exact value) was $10^{-9}$. We include the computed reference matrix $\mathcal E_\mathrm{ref}$ in the supplementary information for this paper~\cite{githubrepo}.

For the computation of the test matrix $\mathcal E_\mathrm{test}$ via the analytical formula~\eqref{eq-surface-i}--\eqref{eq-last}, we tested several numerical integration rules for the triangle: two fixed-order rules from \cite{ecf} with orders 3 and 10 (4 and 25 points respectively), two families of symmetric rules~\cite{Wandzurat2003,Xiao2010} with varying number of points, and also as a baseline the repeated 1d Gauss rule (i.e. by applying the 1d Gauss rule to each of the 2 dimensions of the triangle).

The results are shown in Table~\ref{results-table}, ordered by decreasing relative error $\eta$. In general, for a given number of points all rules displayed approximately the same order of accuracy, for example for each of the 4 rules with 25 points the relative error $\eta$ was $\sim 5 \cdot 10^{-5}$, with the symmetric rule~\cite{Wandzurat2003} showing slightly better accuracy (especially for the 175 point rule with error $4.0 \cdot 10^{-6}$ vs $1.1 \cdot 10^{-5}$ for the symmetric rule~\cite{Xiao2010}). The most accurate rule considered was the repeated 1d Gauss rule with 6400 points; the number of points is clearly too high to use it in practice, but it does show excellent agreement with the reference result~$\mathcal E_\mathrm{ref}$ obtained from finite difference simulations.     

\begin{table}[htb]
\begin{tabular}{lccl}
Method & No. of points $n$ & Rel. error $\eta$ & Ref \\
\hline
Fixed order k=3 &     4  &  $3.3 \cdot 10^{-2}$ & \cite{ecf} \\
Symmetric family 2, k=5 &     7  &  $4.9 \cdot 10^{-3}$ & \cite{Xiao2010} \\
Symmetric family 1, k=5 &     7  &  $4.9 \cdot 10^{-3}$ & \cite{Wandzurat2003} \\
Fixed order k=10 &    25  &  $5.8 \cdot 10^{-4}$ & \cite{ecf} \\
  1d Gauss, m=5 &    25  &  $5.5 \cdot 10^{-4}$ &  \\
Symmetric family 2, k=10 &    25  &  $5.1 \cdot 10^{-4}$ & \cite{Xiao2010} \\
Symmetric family 1, k=10 &    25  &  $4.7 \cdot 10^{-4}$ & \cite{Wandzurat2003} \\
Symmetric family 2, k=20 &    79  &  $4.2 \cdot 10^{-5}$ & \cite{Xiao2010} \\
 1d Gauss, m=10 &   100  &  $4.1 \cdot 10^{-5}$ &  \\
Symmetric family 1, k=20 &    85  &  $3.4 \cdot 10^{-5}$ & \cite{Wandzurat2003} \\
Symmetric family 2, k=30 &   171  &  $1.1 \cdot 10^{-5}$ & \cite{Xiao2010} \\
Symmetric family 1, k=30 &   175  &  $4.0 \cdot 10^{-6}$ & \cite{Wandzurat2003} \\
Symmetric family 2, k=40 &   295  &  $1.7 \cdot 10^{-6}$ & \cite{Xiao2010} \\
 1d Gauss, m=80 &  6400  &  $1.2 \cdot 10^{-8}$ &  \\
\hline
\end{tabular}%
\caption{Numerical integration error using the analytical formula~\eqref{eq-surface-i}--\eqref{eq-last} with various triangle integration rules. For the rules from \cite{ecf,Wandzurat2003,Xiao2010}, $k$ is the order of approximation; for the repeated 1d Gauss rule, $m$ is the number of points of the 1d rule. The weights and integration points for all integration rules are included in the supplementary information for this paper~\cite{githubrepo}. 
 \label{results-table}}%
\end{table}

\section{Summary}
We presented a method to compute the energy of the magnetostatic interaction between linearly magnetized polyhedrons. The magnetostatic energy integral~\eqref{eq-pairwise} is computed using a hybrid procedure where four out of six integration steps are performed analytically resulting in a nonsingular 2d integral~\eqref{eq-surface-i} which is then computed numerically.  

The method can be used in finite element micromagnetics to compute the demagnetizing energy with a high degree of accuracy (for instance, as a reference value in comparison to fast, less accurate traditional methods such as FEM/BEM). Combined with a suitable long range approximation for the magnetostatic integral~\eqref{eq-pairwise}, it can allow an implementation of energy-based fast multpole method (FMM) or tree-code algorithms for the computation of the demagnetizing field.      

Numerical testing showed excellent agreement between the hybrid computation using the new analytical formula~\eqref{eq-surface-i}--\eqref{eq-last} and the reference finite difference simulation.

We thank Prof.~Ronald Cools for providing access to the Online Encyclopaedia of Cubature Formulas~\cite{ecf,Cools2003,Cools1999,Cools1993}. We acknowledge financial support from the EPSRC Centre for Doctoral Training grant EP/G03690X/1.

 \appendix

 \section{Derivation of $\mathbf I_2$}
 \label{sec:app_i2}
 It is easy to verify that $\nabla \R |\R| = \mathbf{Id}|\R| + \R^{\otimes 2}/|\R|$; using eq.~\eqref{eq:decomp} for the first index of the rank-2 tensor $\nabla \R |\R|$ we get
 \begin{align*}
 \mathbf I_2(\partial\tau; \mathbf r') 
&= \int_{\partial\tau - \mathbf r'} \left[
	 \nabla \R |\R| - \mathbf{Id} |\R|
\right] \,ds \\
&= \int_{\partial\tau - \mathbf r'} \left[
		-\n \times (\n \times \nabla \R |\R|)
		+ \n \otimes (\n \cdot \nabla \R |\R|)
		- \mathbf{Id} |\R|
\right] \,ds \\
&= \int_{\partial\tau - \mathbf r'} \left[
        -\n \times (\n \times \nabla \R |\R|)
		+ (\n^{\otimes 2} - \mathbf{Id}) |\R|
		+ (\n \cdot \R) \, \n \otimes \frac{\R}{|\R|}
\right] \,ds
\end{align*}
In the expression $\n \times (\n \times \nabla \R |\R|)$ the cross product acts on the first index of the tensor $\nabla \R |\R|$.

After applying the Stokes' theorem~\eqref{eq:stokes} and noticing that $\n \cdot \R$ is constant over the facets of a polyhedron, we get the desired formula 
\begin{equation}
\mathbf I_2 = 
\sum_{F\in\partial\tau - \mathbf r'}\left[ 
	- \sum_{(\R_1, \R_2) \in \partial F} (\n \times \mathbf u) \otimes \lambda_1(\R_1, \R_2)  
	+ (\n^2 - \mathbf{Id}) \, J_0(F)
	+ (\n \cdot \R_f) \, \n \otimes \mathbf I_1(F)
\right]    
\end{equation}

 \section{Derivation of $\boldsymbol\eta_1$}
 \label{sec:app_eta1}
 To compute $\eta_1(\R_1, \R_2)$, we decompose $\R/|\R|$ into components parallel and orthogonal to $\mathbf u$~(eq.~\eqref{eq:decomp})
 \begin{equation}
\frac{\R}{|\R|} = \nabla |\R| = \mathbf u (\mathbf u \cdot \nabla) |\R| - \mathbf u \times (\mathbf u \times \frac{\R}{|\R|}) 
 \end{equation}
 Integrating over the line $(\R_1, \R_2)$ and using the fact that for points on the line $\mathbf u \times \R$ = $\mathbf u \times \R_1$ 
\begin{equation}
\eta_1(\R_1, \R_2) = 
\mathbf u \int_{\R_1}^{\R_2} (\mathbf u \cdot \nabla) |\R| \, dl 
- \mathbf u \times (\mathbf u \times \R_1) \int_{\R_1}^{\R_2} \frac{1}{|\R|} \, dl 
\end{equation}
 After applying the gradient theorem~(eq.~\eqref{eq:gradient}) we get the desired equation  
 \begin{equation}
 \eta_1(\R_1, \R_2) = 
 	\mathbf u|\R| \, \Big|_{\R_1}^{\R_2}
 	- \mathbf u \times (\mathbf u \times \R_1) \, \eta_0(\R_1, \R_2) 
 \end{equation}

 \section{Derivation of $\boldsymbol\lambda_1$}
 \label{sec:app_lambda1}
We begin by writing the following identities that can be verified by direct differentiation
\begin{align}
(\mathbf u \cdot \nabla) [(\mathbf u \cdot \R) \R |\R|] &=
	\R |\R| 
	+ (\mathbf u \cdot \R) \mathbf u |\R|
	+ (\mathbf u \cdot \R)^2 \frac{\R}{|\R|} \\
(\mathbf u \cdot \nabla) [(\mathbf u \cdot \R)^2 |\R|] &=
	2 (\mathbf u \cdot \R) |\R|
	+ (\mathbf u \cdot \R)^3 \frac{1}{|\R|}
\end{align}
Multiplying the second equation by $\mathbf u/2$ and subtracting from the first to eliminate $(\mathbf u \cdot \R) |\R|$ on the right-hand side, we get
\begin{equation}
(\mathbf u \cdot \nabla) [(\mathbf u \cdot \R) \R |\R|] - 
\frac{\mathbf u}{2} (\mathbf u \cdot \nabla) [\mathbf (\mathbf u \cdot \R)^2 |\R|]
=
\R |\R| + (\mathbf u \cdot \R)^2 \frac{\R}{|\R|} - \frac{\mathbf u}{2} (\mathbf u \cdot \R)^3 \frac{1}{|\R|}
\end{equation}
Solving for $\R |\R|$ and integrating 
\begin{equation}
\boldsymbol\lambda_1(\R_1, \R_2) = 
	\int_{\R_1}^{\R_2} (\mathbf u \cdot \nabla) \big[
		(\mathbf u \cdot \R) (\R  - \frac{\mathbf u}{2} (\mathbf u \cdot \R)) |\R|  
	\big] \,dl
	- \int_{\R_1}^{\R_2} (\mathbf u \cdot \R)^2 [\mathbf{Id}  - \frac{\mathbf u}{2} (\mathbf u \cdot{})] \frac{\R}{|\R|} \,dl
\end{equation}
The first integral can be evaluated using the gradient theorem; to evaluate the second, we note that 
\begin{equation}
(\mathbf u \cdot \R)^2 = |\R|^2  - |\mathbf u \times \R|^2
\end{equation}
Again, the quantity $\mathbf u \times \R$ is constant and can be moved outside of the integral, resulting in the following equation for $\boldsymbol\lambda_1(\R_1, \R_2)$:
\begin{equation}
\boldsymbol\lambda_1 =
	(\mathbf u \cdot \R) (\R  - \frac{\mathbf u}{2} (\mathbf u \cdot \R)) |\R| \, \Big|_{\R_1}^{\R_2}
	- [\mathbf{Id}  - \frac{\mathbf u}{2} (\mathbf u \cdot{})] \boldsymbol\lambda_1 
	+[\mathbf{Id}  - \frac{\mathbf u}{2} (\mathbf u \cdot{})] |\mathbf u \times \R_1|^2 \boldsymbol\eta_1
\end{equation}
In this equation $\boldsymbol\lambda_1(\R_1, \R_2)$ appears on both sides; to solve for it we need to invert the matrix $2\,\mathbf{Id}  - \frac12 \mathbf u (\mathbf u \cdot{})$. It is straightforward to verify that
$(2\mathbf{Id}  - \frac{\mathbf u}{2} (\mathbf u \cdot{}))^{-1} = \frac12\mathbf{Id} + \frac16 \mathbf u (\mathbf u \cdot{})$ and therefore we obtain the desired formula
 \begin{equation}
 \boldsymbol\lambda_1(\R_1, \R_2) =
\big(\frac12\mathbf{Id} + \frac16 \mathbf u (\mathbf u \cdot{})\big)
 \Big[
	(\mathbf u \cdot \R) (\R  - \frac{\mathbf u}{2} (\mathbf u \cdot \R)) |\R| \, \Big|_{\R_1}^{\R_2}
	+[\mathbf{Id}  - \frac{\mathbf u}{2} (\mathbf u \cdot{})] |\mathbf u \times \R_1|^2 \boldsymbol\eta_1(\R_1, \R_2)
 \Big]
  \end{equation}

\section{Vector calculus identities}
\label{sec:app_calculus}
Gradient theorem:
\begin{equation}
\label{eq:gradient}
\int_{\R_1}^{\R_2} (\mathbf u \cdot \nabla) f(\R) \, dl = f(\R_2) - f(\R_1) 
\end{equation}

Decomposition of a vector into components parallel and orthogonal to a unit vector $\mathbf u$ with $|\mathbf u| = 1$: 
\begin{equation}
\label{eq:decomp}
\mathbf a = \mathbf u (\mathbf u \cdot \mathbf a) - \mathbf u \times (\mathbf u \times \mathbf a) 
\end{equation} 

Stokes' theorem (alternative form):
\begin{equation}
\label{eq:stokes}
\int_S (\n_s \times (\n_s \times \nabla)) \, a(\R) \, ds =  \int_{\partial S} a(\R)\mathbf u_l \, dl 
\end{equation}
 
\bibliographystyle{elsarticle-num}

\end{document}